Nanotechnology has rekindled interest in the Kondo effect,
one of the most widely studied phenomena in condensed-matter physics

# Revival of the Kondo effect

### Leo Kouwenhoven and Leonid Glazman

WHY would anyone still want to study a physical phenomenon that was discovered in the 1930s, explained in the 1960s and has been the subject of numerous reviews since the 1970s? Although the Kondo effect is a well known and widely studied phenomenon in condensed-matter physics, it continues to capture the imagination of experimentalists and theorists alike.

The effect arises from the interactions between a single magnetic atom, such as cobalt, and the many electrons in an otherwise non-magnetic metal. Such an impurity typically has an intrinsic angular momentum or "spin" that interacts with the electrons. As a result, the mathematical description of the system is a difficult many-body problem.

However, the Kondo problem is well defined, making it an attractive testing ground for the new numerical and analytical tools that have been developed to attack other challenging many-body problems. Interest in the Kondo effect has therefore persisted because it provides clues to understanding the electronic properties of a wide variety of materials where the interactions between electrons are particularly strong, for instance in heavy-fermion materials and high-temperature superconductors.

Physicists' fascination with the phenomenon has continued since it was first explained by Japanese theorist Jun Kondo in 1964. In fact, interest in the Kondo effect has recently peaked thanks to new experimental techniques from the rapidly developing field of nanotechnology, which have given us unprecedented control over Kondo systems.

### Electron transport at low temperatures

The electrical resistance of a pure metal usually drops as its temperature is lowered because electrons can travel through a metallic crystal more easily when the vibrations of the atoms are small. However, the resistance saturates as the temperature is lowered below about 10 K due to static defects in the material (figure 1a).

Some metals – for example lead, niobium and aluminium – can suddenly lose all their resistance to electrical current and become superconducting. This phase transition from a conducting to a superconducting state occurs at a so-called critical temperature, below which the electrons

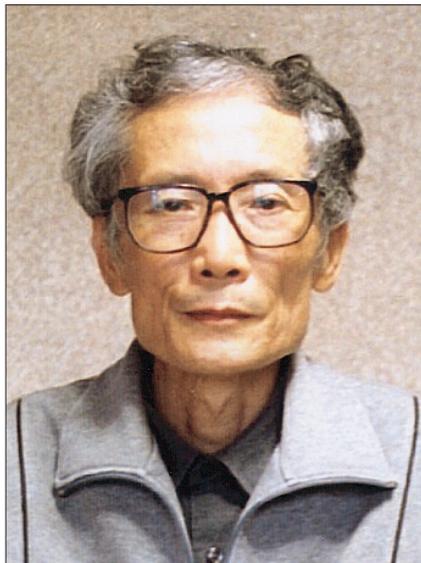

The theory that describes the scattering of electrons from a localized magnetic impurity was initiated by the work of Jun Kondo in 1964

behave as a single entity. Indeed, superconductivity is a prime example of a many-electron phenomenon.

Other metals, like copper and gold, remain conducting and have a constant finite resistance, even at the lowest accessible temperatures. The value of the low-temperature resistance depends on the number of defects in the material. Adding defects increases the value of this "saturation resistance" but the character of the temperature dependence remains the same.

However, this behaviour changes dramatically when magnetic atoms, such as cobalt, are added. Rather than saturating, the electrical resistance increases as the temperature is lowered further. Although this behaviour does not involve a phase transition, the so-called Kondo temperature – roughly speaking the temperature at which the resistance starts to increase again – completely determines the low-temperature electronic properties of the material. Since the 1930s there have been many observations of an anomalous increase in the resistance of metals at low temperature. Yet it took until 1964 for a satisfactory explanation to appear.

Electrical resistance is related to the amount of back scattering from defects, which hinders the motion of the electrons through the crystal. Theorists can readily calculate the probability with which an electron will be scattered when the defect is small. However, for larger defects, the calculation can only be performed using perturbation theory – an iterative process in which the answer is usually written as a series of smaller and smaller terms. In 1964 Kondo made a startling discovery when considering the scattering from a magnetic ion that interacts with the spins of the conducting electrons. He found that the second term in the calculation could be much larger than the first. The upshot of this result is that the resistance of a metal increases logarithmically when the temperature is lowered.

Kondo's theory correctly describes the observed upturn of the resistance at low temperatures. However, it also makes the unphysical prediction that the resistance will be infinite at even lower temperatures. It turns out that Kondo's result is correct only above a certain temperature, which became known as the Kondo temperature, $T_K$.

The theoretical framework for understanding the physics

  33

## 1 The Kondo effect in metals and in quantum dots

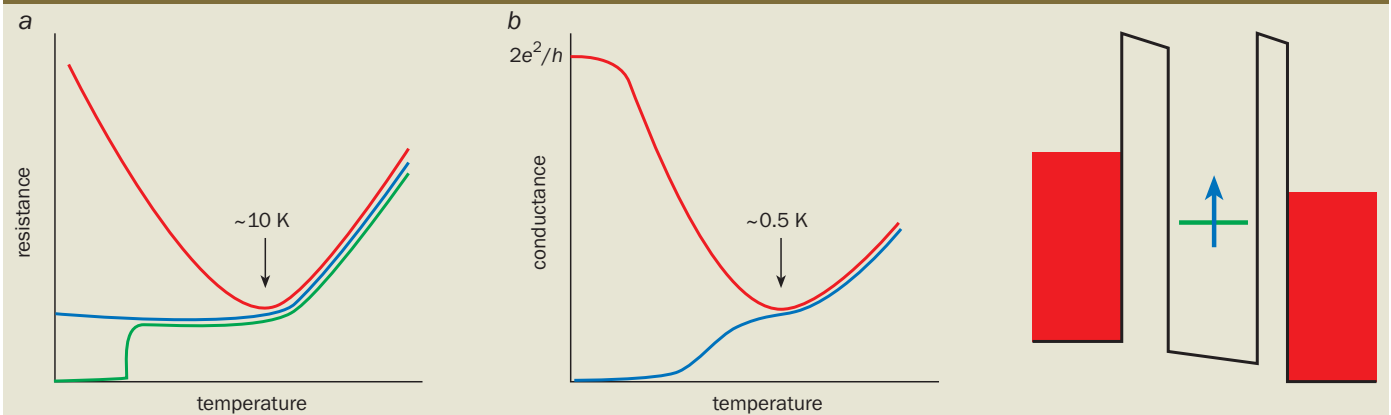

(a) As the temperature of a metal is lowered, its resistance decreases until it saturates at some residual value (blue). Some metals become superconducting at a critical temperature (green). However, in metals that contain a small fraction of magnetic impurities, such as cobalt-in-copper systems, the resistance increases at low temperatures due to the Kondo effect (red). (b) A system that has a localized spin embedded between metal leads can be created artificially in a semiconductor quantum-dot device containing a controllable number of electrons. If the number of electrons confined in the dot is odd, then the conductance measured between the two leads increases due to the Kondo effect at low temperature (red). In contrast, the Kondo effect does not occur when the dot contains an even number of electrons and the total spin adds up to zero. In this case, the conductance continuously decreases with temperature (blue).

below $T_K$ emerged in the late 1960s from Phil Anderson's idea of "scaling" in the Kondo problem. Scaling assumes that the low-temperature properties of a real system are adequately represented by a coarse-grained model. As the temperature is lowered, the model becomes coarser and the number of degrees of freedom it contains is reduced. This approach can be used to predict the properties of a real system close to absolute zero.

Later, in 1974, Kenneth Wilson, who was then at Cornell University in the US, devised a method known as "numerical renormalization" that overcame the shortcomings of conventional perturbation theory, and confirmed the scaling hypothesis. His work proved that at temperatures well below $T_K$, the magnetic moment of the impurity ion is screened entirely by the spins of the electrons in the metal. Roughly speaking, this spin-screening is analogous to the screening of an electric charge inside a metal, although the microscopic processes are very different.

### The role of spin

The Kondo effect only arises when the defects are magnetic – in other words, when the total spin of all the electrons in the impurity atom is non-zero. These electrons coexist with the mobile electrons in the host metal, which behave like a sea that fills the entire sample. In such a Fermi sea, all the states with energies below the so-called Fermi level are occupied, while the higher-energy states are empty.

The simplest model of a magnetic impurity, which was introduced by Anderson in 1961, has only one electron level with energy $\varepsilon_o$. In this case, the electron can quantum-mechanically tunnel from the impurity and escape provided its energy lies above the Fermi level, otherwise it remains trapped. In this picture, the defect has a spin of $\frac{1}{2}$ and its $z$-component is fixed as either "spin up" or "spin down".

However, so-called exchange processes can take place that effectively flip the spin of the impurity from spin up to spin down, or vice versa, while simultaneously creating a spin excitation in the Fermi sea. Figure 2 illustrates what happens when an electron is taken from the localized impurity state and put into an unoccupied energy state at the surface of the Fermi sea. The energy needed for such a process is large, between about 1 and 10 electronvolts for magnetic impurities. Classically, it is forbidden to take an electron from the defect without putting energy into the system. In quantum mechanics, however, the Heisenberg uncertainty principle allows such a configuration to exist for a very short time – around $h/|\varepsilon_o|$, where $h$ is the Planck constant. Within this timescale, another electron must tunnel from the Fermi sea back towards the impurity. However, the spin of this electron may point in the opposite direction. In other words, the initial and final states of the impurity can have different spins.

This spin exchange qualitatively changes the energy spectrum of the system (figure 2c). When many such processes are taken together, one finds that a new state – known as the Kondo resonance – is generated with exactly the same energy as the Fermi level.

The low-temperature increase in resistance was the first hint of the existence of the new state. Such a resonance is very effective at scattering electrons with energies close to the Fermi level. Since the same electrons are responsible for the low-temperature conductivity of a metal, the strong scattering contributes greatly to the resistance.

The Kondo resonance is unusual. Energy eigenstates usually correspond to waves for which an integer number of half wavelengths fits precisely inside a quantum box, or around the orbital of an atom. In contrast, the Kondo state is generated by exchange processes between a localized electron and free-electron states. Since many electrons need to be involved, the Kondo effect is a many-body phenomenon.

It is important to note that that the Kondo state is always "on resonance" since it is fixed to the Fermi energy. Even though the system may start with an energy, $\varepsilon_o$, that is very far away from the Fermi energy, the Kondo effect alters the energy of the system so that it is always on resonance. The only requirement for the effect to occur is that the metal is cooled to sufficiently low temperatures below the Kondo temperature $T_K$.

Back in 1978 Duncan Haldane, now at Princeton University in the US, showed that $T_K$ was related to the parameters of the Anderson model by $T_K = \frac{1}{2}(\Gamma U)^{1/2} \exp[\pi\varepsilon_o(\varepsilon_o + U)/\Gamma U]$, where $\Gamma$ is the width of the impurity's energy level, which

## 2 Spin flips

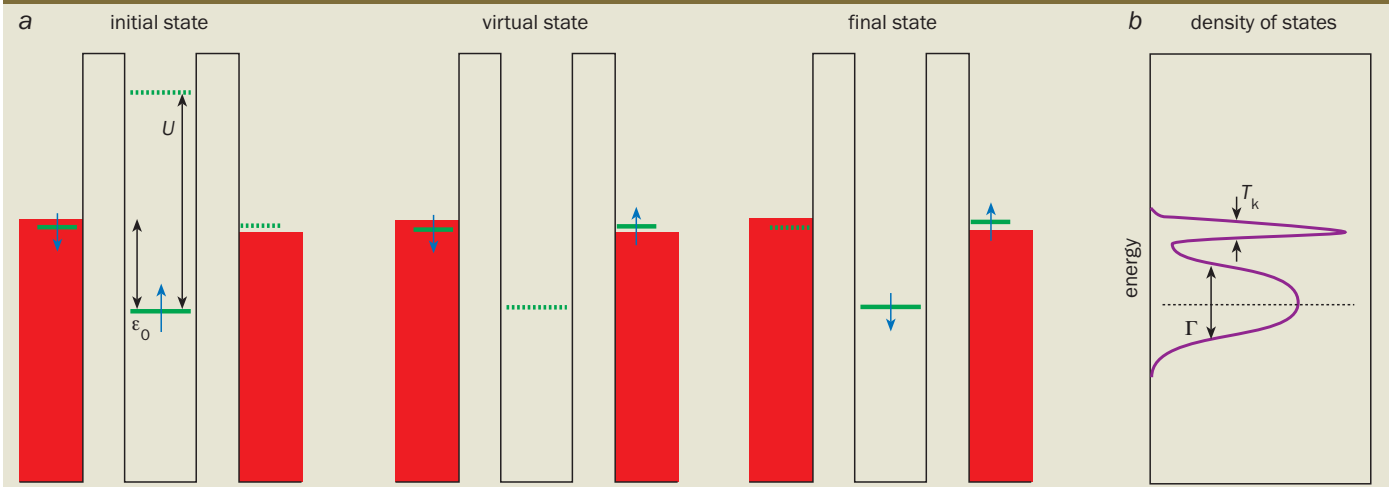

(a) The Anderson model of a magnetic impurity assumes that it has just one electron level with energy $\varepsilon_0$ below the Fermi energy of the metal (red). This level is occupied by one spin-up electron (blue). Adding another electron is prohibited by the Coulomb energy, $U$, while it would cost at least $|\varepsilon_0|$ to remove the electron. Being a quantum particle, the spin-up electron may tunnel out of the impurity site to briefly occupy a classically forbidden "virtual state" outside the impurity, and then be replaced by an electron from the metal. This can effectively "flip" the spin of the impurity. (b) Many such events combine to produce the Kondo effect, which leads to the appearance of an extra resonance at the Fermi energy. Since transport properties, such as conductance, are determined by electrons with energies close to the Fermi level, the extra resonance can dramatically change the conductance.

is broadened by electrons tunnelling from it, and $U$ is the Coulomb repulsion energy between two electrons at the site of the impurity. Due to the exponential dependence on the parameters, the Kondo temperature can vary, in practice, from 1–100 K.

Remarkably, the ratio of the resistance, $R$, divided by the value at absolute zero, $R_0$, depends only on the temperature divided by the Kondo temperature, i.e. $R/R_0 = f(T/T_K)$. Moreover, all materials that contain spin-½ impurities can be described by the same temperature-dependent function, $f(T/T_K)$. So the parameters that characterize the system – $U$, $\varepsilon_o$ and $\Gamma$ – can be replaced by a single parameter, $T_K$.

### Scanning tunnelling microscopy
Nanotechnology aims to manipulate and control matter at the atomic scale. One of the central tools in the field is the scanning tunnelling microscope (STM), which can image a surface with atomic resolution, move individual atoms across a surface and measure the energy spectrum at particular locations.

Recently, the STM has been used to image and manipulate magnetic impurities on the surface of metals, opening a new avenue of research into the Kondo effect. Previously, physicists could only infer the role of the Kondo effect from measurements of resistance and magnetic susceptibility. With the advent of the STM, however, physicists can now simply "photograph" the surface and thereby resolve the position of the atoms prior to studying the phenomenon.

The first results came simultaneously in 1998 from Mike Crommie and colleagues, then at Boston University in the US, and from Wolf-Dieter Schneider's group at the University of Lausanne in Switzerland. Both groups used an STM to image a particular magnetic atom and then to measure the Kondo resonance from the current-versus-voltage characteristics.

More recently, a group led by Don Eigler at IBM's Almaden Research Center in California has built an ellipse of atoms around a cobalt impurity, which was placed at one of the two focal points of the ellipse (see figure 3a). Next, they used an STM to measure the energy spectrum of the cobalt impurity and found a large peak that corresponded to the Kondo resonance. The symmetry of an ellipse is such that electron waves passing through one focus inevitably converge at the second one, thus creating a mirror image of the Kondo resonance. The energy spectrum measured at the second focus also has a Kondo-like peak, in spite of the fact that there is no magnetic impurity at that point. The IBM team has referred to this seemingly unreal situation as a quantum mirage (see Manoharan et al. in further reading).

Meanwhile, Crommie, now at the University of California at Berkeley, and co-workers have moved two magnetic impurities towards each other and studied the interaction between them as a function of their separation (figure 3b).

Scanning tunnel microscopy has moved the Kondo revival in the direction of atom imaging and manipulation, as well as spatially dependent spectroscopy. What an STM cannot do – at least not yet – is alter the properties of the magnetic impurity and its coupling to the metal. In other words, these experiments cannot continually transform one type of magnetic impurity into another with different characteristics. However, this is precisely the direction in which physicists studying quantum-dot devices are moving.

### Quantum dots as artificial magnetic elements
Various groups around the world have exploited chip technology to fabricate small semiconductor devices for investigating fundamental problems in physics. One such device is the quantum dot – a little semiconductor box that can hold a small number of electrons (see Kouwenhoven and Marcus in further reading). Quantum dots are often called artificial atoms since their electronic properties resemble those of real atoms.

A voltage applied to one of the gate electrodes of the device controls the number of electrons, $N$, that are confined in the dot (figure 4). If an odd number of electrons is trapped within the dot, the total spin of the dot, $S$, is necessarily non-zero and has a minimum value of $S = ½$. This localized spin, embedded between large electron seas in the two leads, mimics the cobalt-in-copper system. And many of the known Kondo

## 3 Single magnetic impurities under the microscope

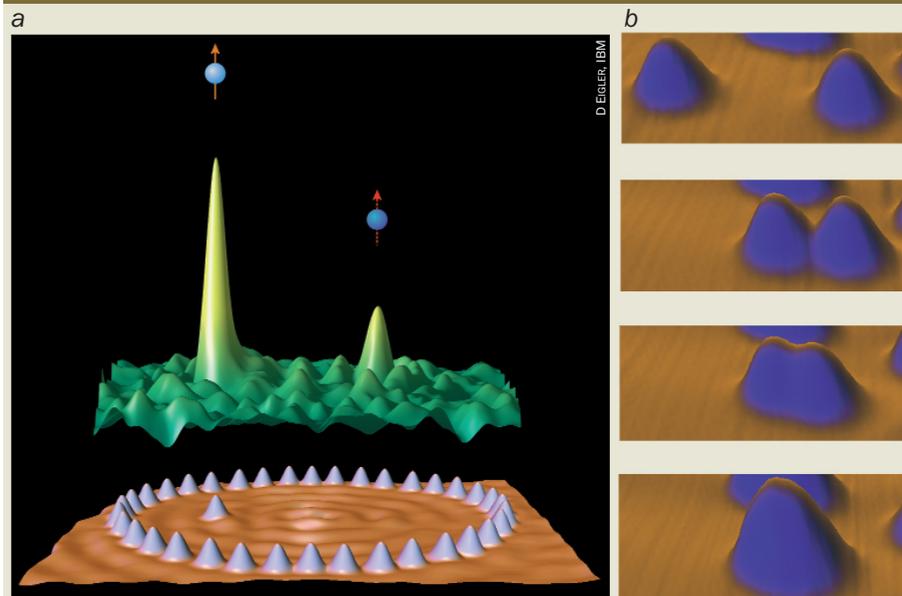

(*a*) By manipulating cobalt atoms on a copper surface, Don Eigler and colleagues at IBM have placed a single cobalt atom at the focal point of an ellipse built from other cobalt atoms (bottom). The density of states (top) measured at this focus reveals the Kondo resonance (left peak). However, elliptical confinement also gives rise to a second smaller Kondo resonance at the other focal point (right) even though there is no cobalt atom there. (*b*) Meanwhile, Mike Crommie and co-workers have measured two Kondo resonances produced by two separate cobalt atoms on a gold surface (top). When two cobalt atoms are moved close together using an STM, the mutual interaction between them causes the Kondo effect to vanish (data not shown).

phenomena can be expected to occur in these transistor-type devices, as was pointed out back in 1988 (see Glazman and Raikh, and Ng and Lee in further reading).

One of the main distinctions between a quantum dot and a real metal is related to their different geometries. In a metal, the electron states are plane waves, and scattering from impurities in the metal mixes electron waves with different momenta. This momentum transfer increases the resistance.

In a quantum dot, however, all the electrons have to travel through the device, as there is no electrical path around it. In this case, the Kondo resonance makes it easier for states belonging to the two opposite electrodes to mix. This mixing increases the conductance (i.e. decreases the resistance). In other words, the Kondo effect produces the opposite behaviour in a quantum dot to that of a bulk metal.

The advantage of quantum dots is the ease with which the parameters of these artificial atoms can be controlled. External "knobs" allow the discrete energy-level structure of the device to be varied, as well as the number of electrons trapped within the dot. In terms of the Anderson impurity model, the energy, $\varepsilon_o$, of the single electron level, its width, $\Gamma$, and the Coulomb repulsion energy, $U$, can all be varied by simply adjusting the voltages on the gates.

Like the resistance of a bulk sample in the Kondo regime, the conductance of a quantum dot depends only on $T/T_K$. With quantum dots, this universality can be readily checked, because the parameters that define $T_K$ can be easily changed with the turn of a knob.

These remarks can be illustrated by some recent results obtained by one of us (LK) and collaborators at Delft University in the Netherlands, NTT in Japan and Tokyo University. Similar experiments have previously been carried out by David Goldhaber-Gordon and co-workers at the Massachusetts Institute of Technology in collaboration with researchers at the Weizmann Institute of Science in Israel (see Goldhaber-Gordon *et al*. in further reading).

At Delft, the conductance of the device was measured as a function of the gate voltage, which changes the number of electrons confined within the quantum dot (figure 5). For an even number of electrons, the conductance decreases as the temperature is lowered from 1 K to 25 mK. This behaviour indicates that the Kondo effect disappears when the number of electrons is even. In contrast, when there is an odd number of electrons, the Kondo effect produces the opposite behaviour, i.e. the conductance increases at low temperatures. Moreover, at the lowest temperatures, the conductance approaches the quantum limit of conductance $2e^2/h$, where $e$ is the charge of an electron.

To analyse these data, we concentrated on the region with $N+1$ electrons and plotted the conductance as a function of temperature for three different values of gate voltage, i.e. three different values of $\varepsilon_o$ (figure 5*b*). The temperature dependence of the conductance is clearly different for the various values of $\varepsilon_o$ although the behaviour of the conductance is similar in each case. The precise temperature dependence was fitted to a function with $T_K$ as a free parameter, which allowed the conductance to be replotted as a function of $T/T_K$ for different values of $\varepsilon_o$ (figure 5*c*). In this so-called normalization plot, the different curves all lie on top of each other, i.e. the data exhibit universal scaling below $T_K$.

The low-temperature increase in conductance and the saturation at $2e^2/h$ are in some sense strange, even though the behaviour is in complete agreement with theory. The system initially contained two potential barriers and a large energy scale, $U$, which tries to block electrons from tunnelling into or out of the dot. Also, the energy $\varepsilon_o$ is far from the Fermi level, i.e. the system is "off resonance". As a result, the set-up is highly unfavourable for electron transport.

However, the "higher-order" spin-flip processes that lead to the Kondo effect completely turn the situation around and increase the conductance until it reaches its ultimate limit. Indeed, the fact that the conductance reaches $2e^2/h$ implies that the electrons are transmitted perfectly through the dot – somehow the Kondo effect is able to make the dot completely transparent.

### Artificial atoms: going beyond real atoms

Quantum dots have provided new opportunities to control the Kondo effect experimentally. Yet in many ways the results described so far – for systems having an odd number of electrons and a spin of ½ – are similar to the old cobalt-in-copper systems. However, quantum dots can also push research into the Kondo effect in new directions, where artificial structures can be exploited in regimes that are inaccessible with magnetic impurities.

The Kondo effect can also occur for impurities and quantum dots that have a spin of 1, or higher. While the spin of an atom is determined by the electronic structure, the spin of quantum dots can be altered more easily. The spectrum of energy levels in a quantum dot can be changed by, for example, applying a magnetic field of around 1 tesla to force a transition between a singlet ($S = 0$) and a triplet ($S = 1$) state. Although the same transition can occur in real atoms, it requires a magnetic field of about $10^6$ tesla, which cannot be generated in the lab.

Exactly at the singlet–triplet transition, we find that several states with different spin and orbital characteristics have the same energy, i.e. the states are said to be degenerate. Exchange processes, like those in figure 2, mix the degenerate states, including those with different orbital angular momenta. The new many-body effect that arises from the degeneracy and exchange interactions is similar to the conventional Kondo effect in many respects. An important difference, however, is that the magnetic field facilitates the effect, rather than destroys it.

Recently, the conductance anomalies associated with the singlet–triplet transition were observed in two very different types of devices (see Pustilnik *et al.* in further reading). First they were found in a rectangular quantum dot made from a semiconductor (figure 4*c*). More recently, Poul Erik Lindelof and co-workers at the University of Copenhagen have observed these conductance anomalies in a molecular electronic device consisting of a carbon nanotube – a thin rolled-up sheet of graphite just a few nanometres in diameter (see Nygård *et al.* in further reading). And Charles Lieber's group at Harvard University has also recently reported the Kondo effect in carbon nanotubes. The measurements were made with an STM near cobalt clusters that were deposited on the nanotubes.

These latest experiments illustrate the generality of the Kondo effect and its importance to nanoelectronic devices. Whenever a small system with a well defined number of electrons is connected to electrodes, Kondo physics affects the low-temperature electronic properties of the device.

Nanotechnology allows physicists to engineer an artificial atom and also design its environment. For example, it is possible to insert a quantum dot into the arm of an electron interferometer (see figure 4*b*). Such ring-shaped devices were pioneered by Moty Heiblum and co-workers at the Weizmann Institute. These two-slit devices enable one to split an electron wave and measure the resulting interference pattern at the point where the two arms reconnect.

For a device with a quantum dot in one arm, the wavefunction of an incoming electron is split into two parts, one part travels through the arm without the dot, while other has to traverse the quantum dot where it experiences the spin interactions of the Kondo effect. Does this major difference in the two paths destroy the interference pattern?

The answer should be no. Unlike a detector, the Kondo effect does not act as an observer who can pass on information about the path taken by the electron; the quantum dot is an integral part of the larger quantum-mechanical system. As long as no one interferes, the interference pattern is preserved. Indeed, recent experiments at Delft and the Weizmann Institute indicate that the interference pattern is not destroyed when the Kondo effect is active on only one of the two slits.

### 4 Quantum-dot devices

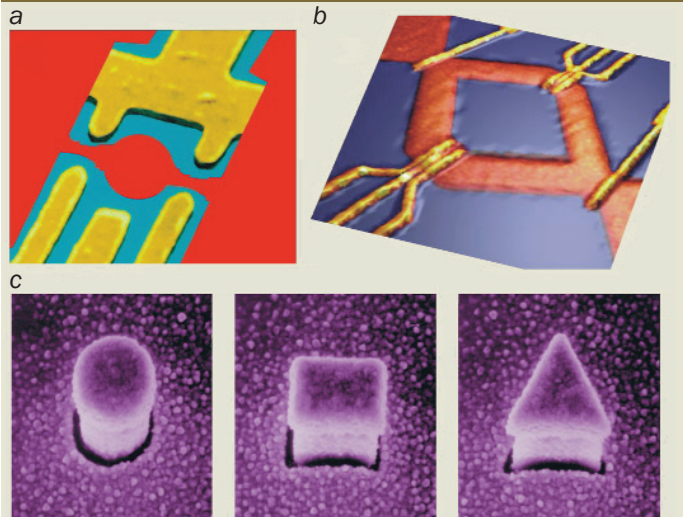

(*a*) A quantum dot can be defined by applying voltages to the surrounding gate electrodes (yellow). The tunnelling between the dot and the external electrodes (top left) is controlled by changing the voltages on the lower-left and lower-right gates. This coupling defines the lifetime broadening, Γ, of the quantum state in the dot. The number of electrons and the energy levels are tuned by the voltage on the lower-central gate. The puddle of electrons (confined red region) is about 0.5 microns in diameter. (*b*) Quantum dots can be placed in both arms of a two-slit interference device. Such a device has been used to investigate whether this scattering destroys the interference pattern. (*c*) Three quantum dots that have been used to compare the Kondo effect for singlet, doublet and triplet spin-states.

### Kondo's future

Scanning tunnel microscopy and quantum-dot devices have provided new tools for studying the Kondo effect from different perspectives and with unprecedented control. Some of the recent studies have counterparts in conventional metal–magnetic-impurity systems, and some are unique to artificial nanostructures.

Investigations into the Kondo effect are far from complete. One ongoing debate concerns the so-called Kondo cloud. The many electrons that are involved in the spin-flip processes in figure 2 combine to build the Kondo resonance. The Kondo cloud consists of electrons that have previously interacted with the same magnetic impurity. Since each of these electrons contains information about the same impurity, they effectively have information about each other. In other words, the electrons are mutually correlated.

The holy grail for research on the Kondo effect is to know whether it is possible to measure and control the Kondo cloud. But perhaps an equally important quest is to understand the time evolution of such a many-body quantum state. For example, how does the state build up? Is it possible to suddenly switch on the exchange interaction in a quantum-dot experiment? Would such experiments allow us to measure how the accompanying Kondo cloud forms?

The Kondo cloud also provides a possible mechanism to investigate the interactions between magnetic impurities. For example, how do the two many-body states that are formed around two separated localized magnetic moments merge? A well controlled study of interacting localized spins could provide us with a new view on extended Kondo systems, such as spin glasses. The basic technology for fabricating interacting Kondo systems now exists, and may soon give birth to yet another Kondo revival.

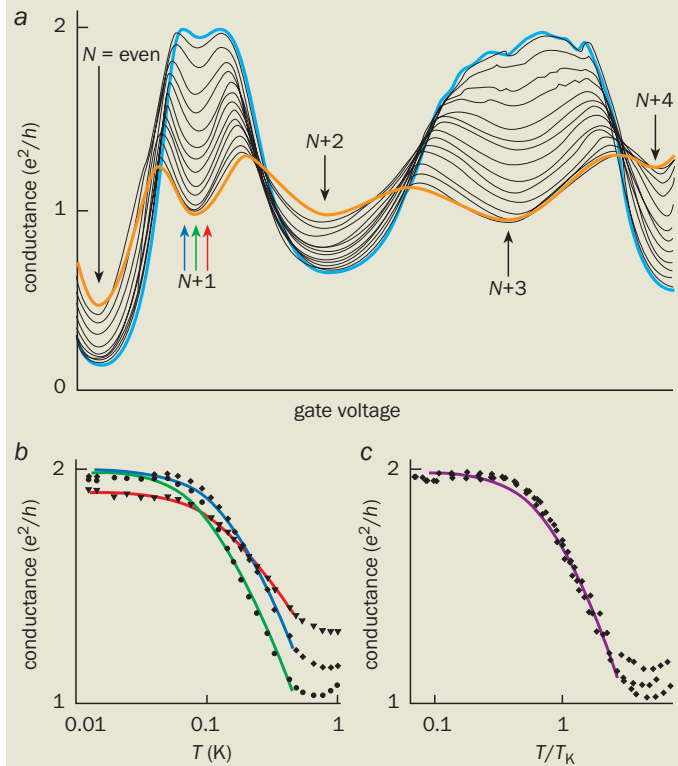

**5 Universal scaling**

(*a*) The conductance (*y*-axis) as a function of the gate voltage, which changes the number of electrons, *N*, confined in a quantum dot. When an even number of electrons is trapped, the conductance decreases as the temperature is lowered from 1 K (orange) to 25 mK (light blue). This behaviour illustrates that there is no Kondo effect when *N* is even. The opposite temperature dependence is observed for an odd number of electrons, i.e. when there is a Kondo effect. (*b*) The conductance for $N + 1$ electrons at three different fixed gate voltages indicated by the coloured arrows in (*a*). The Kondo temperature, $T_K$, for the different gate voltages can be calculated by fitting the theory to the data. (*c*) When the same data are replotted as a function of temperature divided by the respective Kondo temperature, the different curves lie on top of each other, illustrating that electronic transport in the Kondo regime is described by a universal function that depends only on $T/T_K$.

## Further reading

**Leo Kouwenhoven** is in the Department of Applied Physics and the ERATO project on Mesoscopic Correlations, Delft University of Technology, PO Box 5046, 2600-GA Delft, The Netherlands, e-mail leo@qt.tn.tudelft.nl. **Leonid Glazman** is in the Theoretical Physics Institute, University of Minnesota, 116 Church Street SE, Minneapolis MN 55455, USA, e-mail glazman@tpi.umn.edu